\documentclass[aps,prl,twocolumn,nofootinbib]{revtex4-1}
\usepackage{graphicx}
\usepackage{slashed}

\begin{document}

\title{Constraint on parity-violating muonic forces}

\author{Vernon Barger$^1$, Cheng-Wei Chiang$^{2,3,4,1}$, Wai-Yee Keung$^{5,6}$, and
  Danny Marfatia$^{7,1}$}
\vspace{0.5cm}
\affiliation{%
\bigskip
$^1$ Department of Physics, University of Wisconsin, Madison, WI 53706, USA \\
$^2$ Department of Physics and Center for Mathematics and Theoretical Physics,
National Central University, Chungli, Taiwan 32001, ROC \\
$^3$ Institute of Physics, Academia Sinica, Nankang, Taipei 11529, ROC \\
$^4$ Physics Division, National Center for Theoretical Sciences, Hsinchu 30013, ROC \\
$^5$ Department of Physics, 
University of Illinois at Chicago,
Chicago,  IL 60607-7059, USA \\
$^6$ Department of Physics, Brookhaven National Laboratory, 
      Upton, NY 11973, USA \\
$^7$ Department of Physics and Astronomy,
University of Kansas, Lawrence, KS 66045, USA
}

%\author{Vernon Barger}
%\affiliation{Department of Physics, University of Wisconsin-Madison, Madison, WI 53706, USA}

%\author{Cheng-Wei Chiang}
%\affiliation{Department of Physics and Center for Mathematics and Theoretical Physics,
%National Central University, Chungli, Taiwan 32001, ROC}
%\affiliation{Institute of Physics, Academia Sinica, Nankang, Taipei 11529, ROC}
%\affiliation{Physics Division, National Center for Theoretical Sciences, Hsinchu 30013, ROC}
%\affiliation{Department of Physics, University of Wisconsin-Madison, Madison, WI 53706, USA}

%\author{Wai-Yee Keung}
%\affiliation{Department of Physics, University of Illinois at Chicago, Chicago,  IL 60607-7059, USA}
%\affiliation{Department of Physics, Brookhaven National Laboratory, Upton, NY 11973, USA}

%\author{Danny Marfatia}
%\affiliation{Department of Physics and Astronomy, University of Kansas, Lawrence, Kansas 66045, USA}
%\affiliation{Department of Physics, University of Wisconsin-Madison, Madison, WI 53706, USA}

\vskip 2cm
\begin{abstract}
Using the nonobservance of missing mass events in the leptonic kaon decay $K\to \mu X$, we place a strong constraint on exotic parity-violating gauge interactions of the right-handed muon. 
By way of illustration, we apply it to an explanation of the proton size anomaly that invokes such a new force; scenarios
in which the gauge boson decays invisibly or is long-lived are constrained.
\end{abstract}

\maketitle

%\newpage
%%
%\section*{Introduction}

%{\it Introduction ---}
 In the standard model (SM), the right-handed charged lepton field $\ell_R$ is a gauge singlet, and the chiral muon field $\mu_R$ is an example of such a field.  It is straightforward to add a new $U_{\mu_R}(1)$ gauge interaction without modifying the SM gauge group structure, and simultaneously evade many phenomenological constraints.  Recently, this possibility has been entertained~\cite{Batell:2011qq} to explain a measurement of the proton radius obtained from the Lamb shift of muonic hydrogen~\cite{Pohl:2010zz} , that is $5\sigma$ smaller than that determined from ordinary hydrogen or $e$-$p$ scattering data~\cite{Mohr:2008fa}.  
 While the new interaction alone would be in conflict with measurements of 
 the muon anomalous magnetic dipole moment $g_\mu-2$~\cite{Barger:2010aj},  one can arrange a delicate cancellation from another sector of new physics, such as a new scalar boson associated with the Higgs mechanism.  Although unnatural, such fine tuning is conceivable.

An explicit example of such a cancellation can be found in the model of Ref.~\cite{Batell:2011qq} which has a 
$U_{\mu_R}(1)$ vector gauge boson $V$ and a complex scalar field, both with mass of tens of MeV.  
The Lamb shift correction in muonic hydrogen is accounted for by a modest gauge coupling $g_R\approx 0.01$ and a small kinetic mixing amplitude \mbox{$\kappa \sim 0.002$} between $V$ and the photon field.  
The large $V$-exchange contribution to $g_\mu-2$ is cancelled at the $0.1\%$ level by the contribution of the scalar.

In this Letter, we examine an important constraint on the $g_R$ gauge coupling to $\mu_R$ in the context of the leptonic kaon decay, $K \to \mu\nu$~\cite{Pang:1989ut}.  If $V$ is lighter than 100~MeV, it can be radiated from the muon line of the above process.  If $V$ is stable, the combined recoiling system forms a missing mass for which
there is no experimental evidence.  In fact, the size of $g_R$ that accommodates the Lamb shift of 
muonic hydrogen~\cite{Batell:2011qq} is not allowed by leptonic kaon decay provided $V$ decays invisibly or does not decay inside the detector. 

Note that in the minimal version of the model of Ref.~\cite{Batell:2011qq},
$V$ decays promptly into $e^+e^-$ pairs via kinetic mixing with
the photon, and our constraint does not apply.\footnote{
Measurements of $K^+ \to \mu^+ \nu e^+e^-$ have been made with $e^+e^-$ invariant masses above 
145~MeV~\cite{hep-ex/0204006}, so that they are relevant only for $m_V>145$~MeV.

However, a recent search for $V$ in the decay chain $\phi \to \eta V$, \mbox{$\eta\to \pi^+\pi^- \pi^0$}, $V \to e^+ e^-$, by the KLOE-2 collaboration~\cite{kloe} excludes the kinetic mixing parameters corresponding to the points with \mbox{$(m_V, g_R)=(50~\rm{MeV},0.05)$} and $(100~\rm{MeV},0.07)$ in 
Ref.~\cite{Batell:2011qq}. The $(m_V, g_R)=(10~\rm{MeV},0.01)$ point of Ref.~\cite{Batell:2011qq} 
yields a proton-muon interaction that is incompatible with measurements of the muonic 3D$_{5/2}$ -- 2P$_{3/2}$ X-ray transition in $^{24}$Mg and $^{28}$Si~\cite{Beltrami}.  Other points of the minimal scheme that survive these constraints may exist, but this requires a parameter space scan.}  More baroque realizations, in which there are new particles that are charged under $U_{\mu_R}(1)$ and lighter than $m_V/2$, are strongly constrained unless these particles decay to the SM. 

%\section*{A light vector coupled to $\mu_R$}

\begin{figure}[t]
\centering
\includegraphics[width=3in]{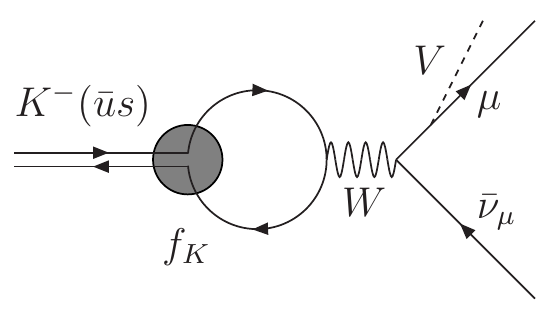}
\caption{\label{fig:zbrems}%
$V$ bremsstrahlung in $K^- \to \mu^- \bar{\nu}_\mu$ decay.}
\end{figure}

%{\it Analysis ---} 
For the sake of generality, we assume that a light vector 
particle $V$ and the right-handed muon interact via the Lagrangian term,
%
%\begin{equation}
%{\cal L} \supset  
\begin{equation}
g_R \bar \mu_R \slashed V \mu_R ~.
\end{equation}
%\end{equation}
%
It is possible to produce a $V$ boson by radiation in \mbox{$K \to \mu \nu$} decay as long as the $V$ boson is lighter than about 100~MeV; see Fig.~\ref{fig:zbrems}.

In the process $K^-\to \mu^- V \bar\nu_\mu$, the relevant hadronic weak-current matrix element is
$\langle 0| \bar u\gamma^\alpha(1-\gamma_5) s|K^-\rangle=f_K p_K^\alpha$, where $p_K^\alpha$ denotes the momentum of the decaying kaon and $f_K = 156.1$ MeV~\cite{PDG}.
The amplitude for the process is then
\begin{eqnarray}
{\cal M}=
%{ G_F\over\sqrt{2}}
%\bar u_\mu g_R \not \epsilon_V \fr{1+\gamma_5}{2} 
%{\not p_\mu +\not p_V +m_\mu \over
%                           (p_\mu+p_V)^2-m_\mu^2} 
%       \sin\theta_C f_K\not K (1-\gamma_5)  v_{\bar \nu} $$
%=
{\sqrt{2} g_R \, G_F \, f_K \, m_\mu \, \sin\theta_C
\over
(p_\mu+p_V)^2-m_\mu^2}
\left[
\bar u_\mu \slashed\epsilon_V \slashed p_K \frac{1-\gamma_5}{2} v_\nu
\right]\,,
\end{eqnarray}
where $\theta_C$ is the Cabibbo angle and $\epsilon^\mu_V$ is the polarization vector of the $V$ boson.  The spin-summed squared amplitude is given by
\begin{eqnarray}
&& \sum |{\cal M}|^2
\nonumber \\
&&= 
\frac{4 g_R^2 G_F^2 f_K^2 m_\mu^2\sin^2\theta_C}
{(m_V^2 +2p_V\cdot p_\mu )^2}
\Bigl[
2 p_K\cdot p_\mu \, p_K\cdot p_\nu - m_K^2 p_\mu\cdot p_\nu
\Bigr. \nonumber \\
&& \quad \Bigl.
+ \frac{2 p_V\cdot p_\mu}{m_V^2}
( 2p_K\cdot p_V \, p_K\cdot p_\nu - m_K^2 p_V\cdot p_\nu)
\Bigr] ~.
\label{ampsquared}
\end{eqnarray}
In the rest frame of the kaon, energy conservation in terms of the scaling variables,
\begin{eqnarray}
x_\alpha &=& 2E_\alpha/m_K = 2p_K\cdot p_\alpha/m_K^2~,~\ \ \ \ \  \alpha = \mu, \nu, V
\nonumber
\end{eqnarray}
dictates $x_\mu + x_\nu + x_V=2$. We have for the scalar products,
\begin{eqnarray}
p_\mu\cdot p_\nu &=& \frac{m_K^2}2 (1-x_V+\delta_V-\delta_\mu)  ~, 
\nonumber \\
p_\mu\cdot p_V &=& \frac{m_K^2}2 (1-x_\nu-\delta_V-\delta_\mu) ~, 
\\
p_\nu\cdot p_V &=& \frac{m_K^2}2 (1-x_\mu-\delta_V+\delta_\mu) ~,
\nonumber
\end{eqnarray}
with $\delta_V=m_V^2/m_K^2$ and $\delta_\mu=m_\mu^2/m_K^2$.  We thus derive the differential decay rate
\begin{eqnarray}
\label{eq:diffxsec}
\frac{d\Gamma(K^-\to \mu^- V \bar\nu_\mu)}
{dx_\mu dx_\nu}
= {m_K\over 256\pi^3 } \sum |{\cal M}|^2 ~,
\end{eqnarray}
with $\sum |{\cal M}|^2$ in Eq.~(\ref{ampsquared}) written in terms of $x_{\mu,\nu,V}$ and $\delta_{\mu,V}$.  The range of $x_\mu$ is 
$\left[ 2\sqrt{\delta_\mu}, 1+\delta_\mu-\delta_V \right]$.  $x_\nu$ is bounded by the following upper and lower limits:
\begin{eqnarray}
&& \frac{1}{2(1-x_\mu+\delta_\mu)}
\Bigl[ 
(2-x_\mu)(1-x_\mu+\delta_\mu+\delta_V)
\Big. \nonumber \\
&& \qquad
\Big. \pm \sqrt{x_\mu^2-4\delta_\mu}(1-x_\mu+\delta_\mu-\delta_V)
\Bigr]\,.
\end{eqnarray}
It is useful to normalize our result in Eq.~(\ref{eq:diffxsec}) with respect to the standard two-body decay rate,
\begin{eqnarray}
\Gamma(K^-\to\mu^-\bar\nu_\mu)=\frac{G_F^2}{8\pi} 
m_K m_\mu^2 f_K^2 \sin^2\theta_C 
\left( 1 - \frac{m_\mu^2}{m_K^2} \right)^2
\end{eqnarray}
to get the dimensionless formula
\begin{eqnarray}
&& \frac1{\Gamma(K^-\to \mu^- \bar\nu_\mu)}\frac{d\Gamma(K^-\to \mu^- V \bar\nu_\mu)}{dx_\mu dx_V}
\nonumber \\
&&= \frac{g_R^2 /(1-\delta_\mu)^2}{16\pi^2 (1-\delta_\mu-x_\nu)^2 } 
\Bigl[
x_\mu x_\nu -1 +x_V-\delta_V+\delta_\mu
\Bigr.
\nonumber \\
&& \quad
\Bigl.
+ \frac1{\delta_V} (1-x_\nu-\delta_V-\delta_\mu)
    (x_V x_\nu-1+ x_\mu+\delta_V-\delta_\mu )
\Bigr]  ~.
\nonumber \\
\end{eqnarray}
After integrating over $x_\nu$, the resulting energy distribution in $x_\mu$ can be confronted by the search for a missing recoiling mass in muonic kaon decay. To compare with experiment, we need 
${1\over  \Gamma(K^-\to \mu^- \bar\nu_\mu)} { d\Gamma(K^-\to \mu^- X)\over d m_{X}}$  versus $m_X$, with $X$ denoting the missing energy.  
Since $p_X=p_V + p_\nu$, we get $m_X^2=m_K^2(1-x_\mu+\delta_\mu)$, and
\begin{eqnarray}
\frac{d\Gamma}{d m_X}
= 
\frac{2\sqrt{1-x_\mu+\delta_\mu}}{m_K} 
\frac{d\Gamma}{d x_\mu} ~.
\end{eqnarray}

\begin{figure}[t]
\centering
\includegraphics[width=3.1in]{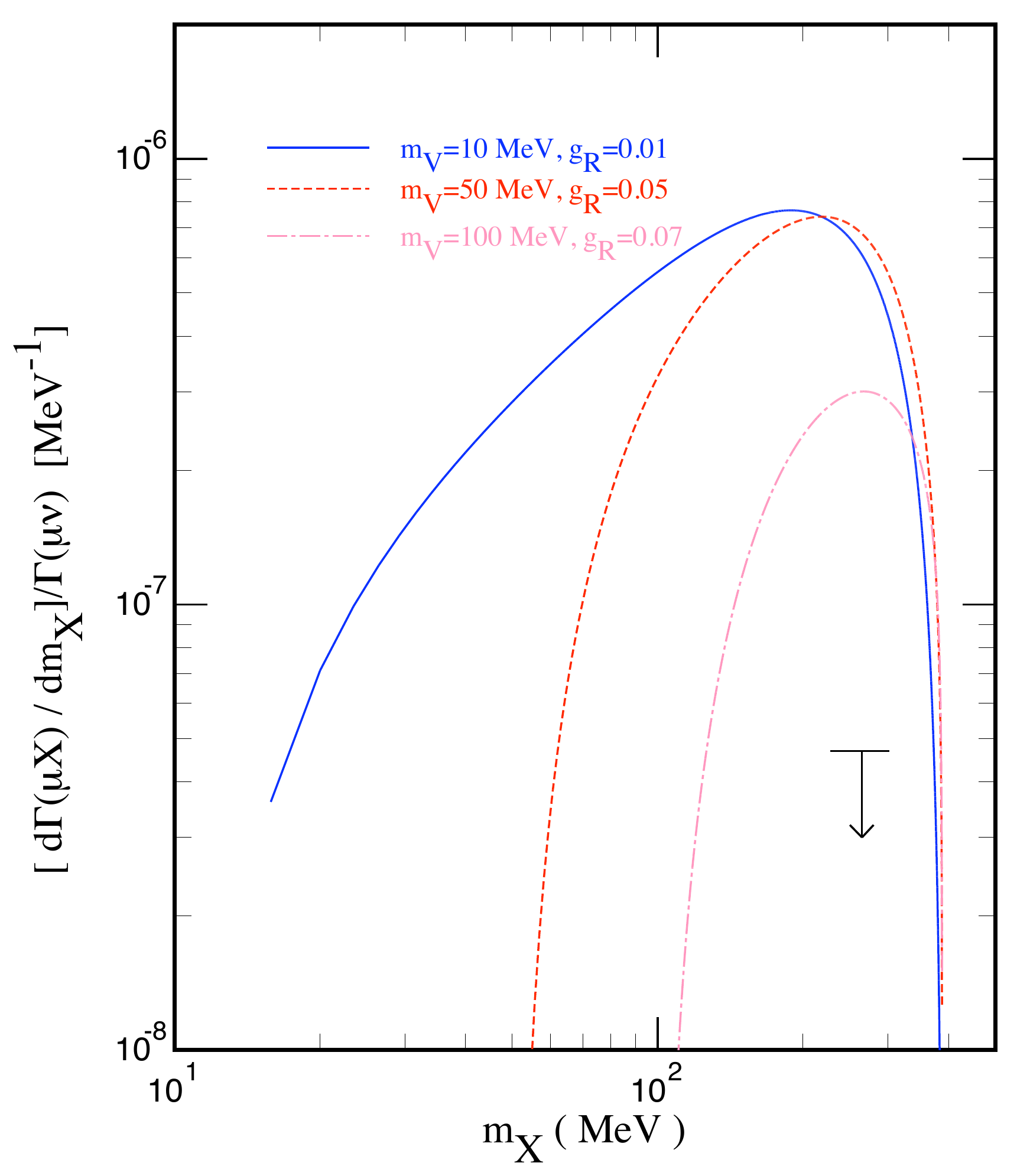}
\caption{\label{fig:events}%
Differential decay rate of muonic kaon decay with $V$ bremsstrahlung as a function of the missing mass, normalized to the standard two-body muonic kaon decay. The 90$\%$ CL upper limit in the mass range 
$227.6 \leq m_X \leq 302.2$~MeV is marked by a short horizontal line.  The distributions for the three benchmark points shown violate the upper limit. We remind the reader that the bound is evaded by the minimal model of 
Ref.~\cite{Batell:2011qq}, since $V$ decays promptly to $e^+e^-$; model extensions in which $V$ decays invisibly
or is long-lived are strongly constrained.  }
\end{figure}

%{\it Constraint ---} 
A null result for missing mass in such decays was obtained with a sensitivity of $10^{-7}$~MeV$^{-1}$~\cite{Pang:1989ut}.  The experimental acceptance of the muon kinetic energy is in the range, 60~MeV to 100~MeV, that corresponds to a missing mass $m_X$ of $302.2$~MeV to $227.6$~MeV, a mass interval of $74.6$ MeV.  The nonobservation of a signal sets a 90\% CL upper limit on the branching fraction of $3.5\times 10^{-6}$ in this mass interval, corresponding to a normalized differential fraction $4.7\times10^{-8}$ MeV$^{-1}$.  In previous work, this limit has been used to constrain the
Majoron model~\cite{Barger:1981vd}.  
%Here we apply this limit to the $V$ bremsstrahlung in the muonic kaon decay.

\begin{figure}[t]
\centering
\includegraphics[width=3in]{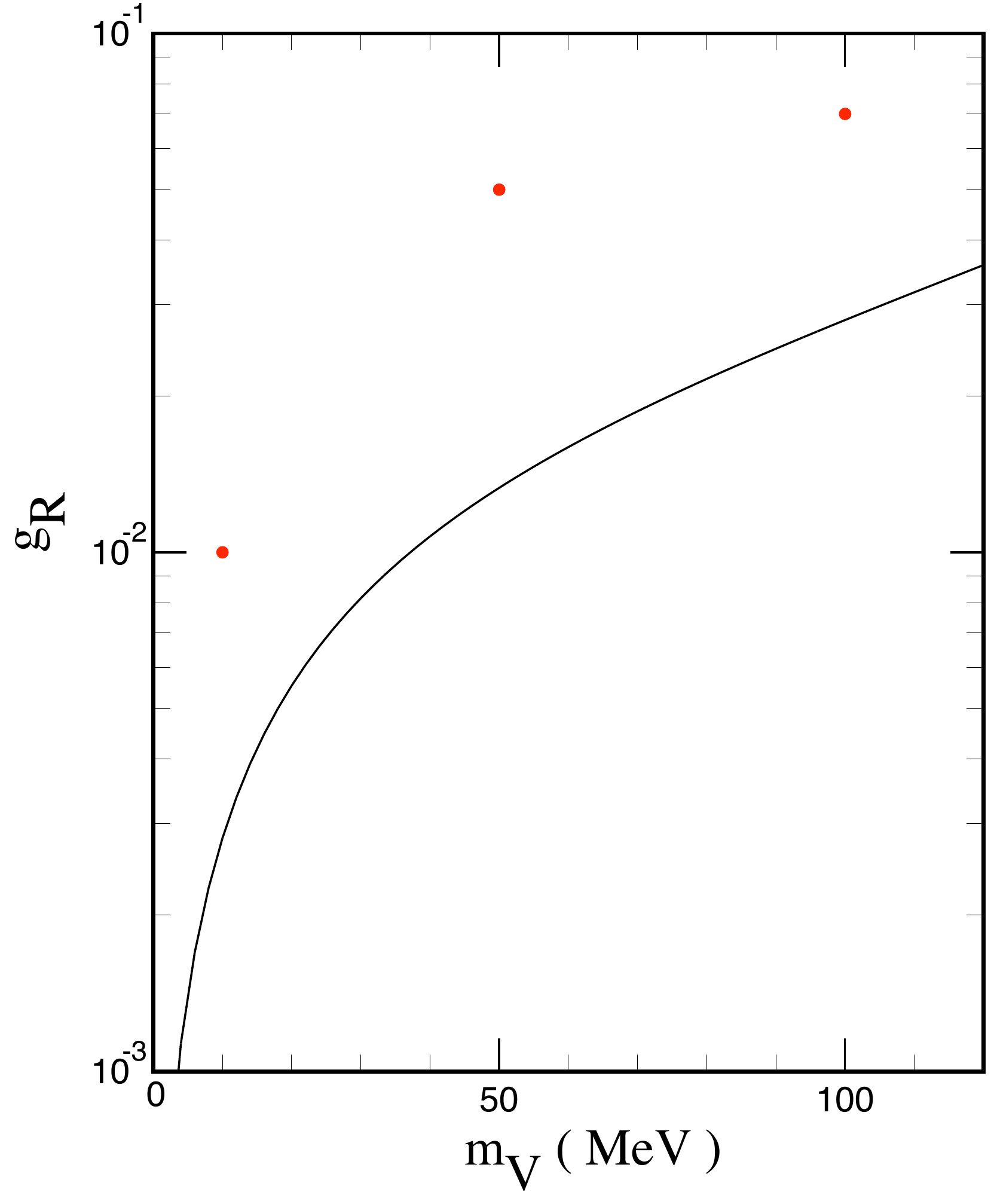}
\caption{\label{fig:coupling}%
The $(m_V,g_R)$ parameter space above the solid curve is excluded at the 90\%~CL. The three red dots are the benchmark points in Fig.~\ref{fig:events} and are disallowed if $V$ decays invisibly
or is long-lived.}
\end{figure}

In Fig.~\ref{fig:events}, we show the normalized differential decay rate of $K \to \mu V \nu$ as a function of the missing mass.  The short horizontal line marks the 90$\%$ confidence level (CL) upper limit in that mass range.  We also show the differential decay rate curves corresponding to three benchmark choices of $(m_V,g_R)$ for the model of 
Ref.~\cite{Batell:2011qq} with the assumption that $V$ has a long enough lifetime that it does not decay inside the detector, or that it decays invisibly.  The 90\% CL upper limit on $g_R$ is shown in Fig.~\ref{fig:coupling}.  The three benchmark choices of 
Fig.~\ref{fig:events} indicated by red dots are disallowed.

%Their event distributions are in conflict with the experimental constraint.  
%We note in passing that in the model of Ref.~\cite{Batell:2011qq} it is also possible to have $\phi$-emission in $K \to \mu \nu$ decay which adds more missing mass events, thus exacerbating the situation. Also, no fine tuning or cancellation can help relax this constraint.

%{\it Conclusion ---} 
In conclusion, we pointed out a constraint on a new gauge interaction that couples to the right-handed muon and has a gauge boson mass less than about 100~MeV.  This light gauge boson 
can be copiously produced by bremsstrahlung off the muon line in $K \to \mu\nu$ decays. The lack of experimental 
evidence for missing mass events constrains the size of the coupling and variants of a model~\cite{Batell:2011qq} proposed to explain the proton size anomaly.

%\section*{Acknowledgement}
{\it Acknowledgments.}
We thank B.~Batell, D.~McKeen and M.~Pospelov for correspondence and W.~Marciano for discussions.  WYK thanks BNL for its hospitality during his visit. DM thanks the University of Hawaii for its hospitality during the completion of this work. This research 
was supported by DoE Grant Nos. DE-FG02-84ER40173, 
DE-FG02-95ER40896 and DE-FG02-04ER41308, by NSF 
Grant No. PHY-0544278, by NSC Grant No. 100-2628-M-008-003-MY4, by NCTS, and by the Wisconsin Alumni Research Foundation.

\end{document}